\documentstyle[prd,aps,preprint]{revtex}
\begin{document}
\draft

\preprint{Imperial/TP/93-94/33 and hep-ph/9405222}


\title{Interactions between $U(1)$ Cosmic Strings: an analytical study}
\author{L. M. A. Bettencourt\footnote{e-mail: l.bettencourt@ic.ac.uk}
 and R. J. Rivers}
\address{The Blackett Laboratory, Imperial College,
London SW7 2BZ, U.K.}
\date{\today}
\maketitle

\begin{abstract}
\noindent We derive analytic expressions for the interaction energy
between  two general $U(1)$ cosmic strings
as the function  of their relative orientation and the ratio of the coupling
constants in the model. The  results are relevant to the
statistic description of strings away from critical coupling and shed
some light on the mechanisms involved in string formation and the
evolution of string networks.
\end{abstract}



\section{Introduction}

Phase transitions associated with the spontaneous breaking of
symmetries in gauge theories are expected to have played a crucial
role in the history of the early universe. In particular they imply the
production of topologically stable defects \cite{Kib}.
Such defects, produced at the energy scale of grand
unification,  lead to a panoply of
cosmological effects. The most important consequence of their
existence, however, is their potential to create energy
distribution anisotropies, necessary for seeding the large scale
structure of the present universe, and compatible with those observed in the
cosmic microwave background \cite{Struc}.

Different classes of defects lead to distinct cosmological
consequences. The most dramatic  is  that walls and
monopoles are known to come to dominate the energy density of the
universe, if inflation does not occur subsequently to their formation
or some other mechanism intervenes to enhance their annihilation.
Strings by contrast, under certain general conditions,  seem to be
entirely viable on their own.
The fundamental difference arises from the fact that a network of
cosmic strings possesses natural mechanisms to reduce its own
contribution to the total energy density. Moreover, this is done in a way such
that the effective evolution of the energy fraction in strings
decreases in time faster than would be expected from the expansion
alone  and scenarios of string domination are thereby naturally precluded.

The mechanisms that allow for the viable evolution of a string network
are essentially motivated on topological grounds and are consequently
expected to be model independent \cite{Shellard,Niel}. They consist
of  a two step process. Firstly, when two segments of a long
string intersect a closed loop of string will be formed. This
happens when the two  segments exchange ends
after collision, which in turn is a consequence of winding number
(i.e., vortex topological charge)
conservation on the plane. This is designated intercommuting.
Finally, because
closed loops of string are not globally topologically stable they can
radiate away the energy trapped in their field configuration  as
gravitational waves and shrink until they disappear. The final stage
of collapse is probably characterized by vortex annihilation and is
expected to produce extremely energetic cosmic rays.

The realization of this scenario rests crucially upon the
efficiency of the intercommuting and subsequent string separation.
Strings, however are know to experience interactions which depend, in
general both on the underlying field theory and on the specific region
of its parameter space.
In its simplest and most widely used model strings are the classical
non-trivial solutions of the Abelian Higgs model. In particular, for
certain ratios of coupling constants,  they
are analogous to the vortex solution in type II superconductors
and to vortices in superfluid $^4 He$. This latter case corresponds
to the vanishing of the gauge sector in the model and the
corresponding string solutions are naturally known as global strings.

Typically, when estimating how string networks evolve, numerical
simulations only invoke the underlying field theory in the initial string
formation. Thereafter, they are taken to obey  the equations of
classical Nambu-Goto strings, interaction-free but for empirical
assumptions about intercommuting. For type II and global strings these
assumptions are buttressed by a number of numerical studies.
They reveal, in particular, that
strings indeed exchange ends and separate in these cases. Studies
regarding the outcome of collisions of type I strings, which feel an
attractive interaction regardless of orientation, are much scarcer
\cite{Lag}.
On the other hand, analytical studies concerning the derivation and
generalization
of such  range of behavior having the field theory as
starting point
are scarce and tend to concentrate on specific regions of
parameter space \cite{She,Bat}. In this paper we attempt a more
general analytic analysis.

The article is organized in two parts. The longer is concerned with
classical vortex interactions. The first sections essentially review
the approximate field solutions for vortices and derive the
interaction potential between two vortices or strings, for all values
of coupling constant ratios and arbitrary orientations. Models for
string production typically lead to initial high string densities.
The latter part of the paper examines the initial stability of string
networks, both classically and quantum mechanically. This part of the
paper is rather more speculative. Type I and type II strings are a
consequence of theories that, essentially, undergo either
first-order or second-order transitions at the time of string
formation. We make a preliminary attempt to relate this to the nature
of string forces.

\section{ The Interaction Energy between two strings: generalities}

Prior to considering multiple string solutions we briefly
review  the isolated string field
solutions, with particular attention to the field behavior
at large distances  from their axes of symmetry. Although well-known
we need the details for later sections.

$U(1)$ strings are the extrapolation to three dimensions of the
well-known vortex solution of the Abelian Higgs model \cite{Nie}.
These are finite energy non-trivial static solutions on the plane. The
Lagrangian density takes the  form
\begin{eqnarray}
{\cal L} = -{1 \over 4 } F_{\mu \nu} F^{\mu \nu} + {1 \over 2} \vert
(\partial_\mu + i e A_\mu) \phi \vert ^2 -{\lambda \over 8} (\vert \phi
\vert ^2 - \eta^2)^2,
\end{eqnarray}
\noindent where $e$ is the gauge coupling constant to the scalar
field, $\lambda$ the
scalar field self-coupling constant and $\eta$ the vacuum expectation
value of the modulus of the scalar field in the broken phase, as usual.

The finiteness of the energy alone determines the asymptotic form of
the fields at infinity. In order for the scalar field  potential to
vanish at infinity, $\phi$ is constrained to lie on a circle of radius
$\eta$. Its magnitude
becomes then well defined but its phase remains arbitrary. On the other
hand, the modulus of the covariant derivative must also vanish
asymptotically which forces the gauge field to assume the form \cite{Jac}
\begin{equation}
lim_{ \vert \vec x \vert \rightarrow + \infty} A_\theta(x) = -
{i \over e} \partial_i \phi(x).
\end{equation}
This also ensures that the energy contribution arising from the field
strength tensor will be finite.
Moreover, if we work in polar coordinates, we see that the problem of
parameterizing the behavior of the scalar field at infinity reduces
to the mapping of a circle, in coordinate space, onto another, in field
space. These mappings, as is well known, fall into an infinite number of
homotopy classes, each labeled by an integer $n \in Z$. Each solution,
characterized by a given $n$, cannot in turn, be continuously deformed
into another with $m\neq n$, without changing the boundary conditions
for the fields at infinity. Since this would imply expending an infinite
amount of energy strings are topologically stable.

Even though the above arguments supply  us with with some information
about vortex solutions the way in which the fields tend to their
asymptotic forms is only revealed by studying the
Euler-Lagrange equations. These can be solved numerically in all
detail but, for
certain purposes, this does not suffice. Such is the case when we want
to discuss analytically the detailed energetics of a multiple string
field configuration.

Exact single solutions exist only for the case of critical coupling ($b =
{e^2 \over \lambda} = 1$) and positive winding number \cite{Vega}.
This is a rather
special case since the interactions between vortices disappear and the
equations of motion for the fields become first order; the well-known
Bogomol'nyi equations.
In general, however, approximate solutions in certain regimes can be
derived. In order to show this we take the Euler-Lagrange
equations, assuming cylindrical symmetry, in the radial temporal
gauge $A_r=A_0=0$
\begin{eqnarray}
\nabla_r^2 \varphi - \left[ \left( e A_\theta - {n \over r } \right)^2 +
{\lambda \over 2 } \left( \varphi^2 - \eta \right) \right]  \varphi &=& 0 \\
\nabla \times  \nabla \times A_\theta   +
 e \varphi^2 \left( e A_\theta - {n \over r} \right) &=& 0,
\end{eqnarray}

\noindent where $\varphi$ is the modulus of the scalar field and $n$ its
winding number. The field $\phi$ is therefore assumed to be of the
form $\phi(r,\theta)=\varphi(r) e^{i n \theta }$, where $\theta$ is the angular
co-ordinate on the plane.
To probe how the fields approach their asymptotic values at infinity,
one can expand $\varphi$ around $\eta$ as $\varphi(r) = \eta - f(r)$,
where $f$ is  an auxiliary field.
Then the Euler-Lagrange equations become

\begin{eqnarray}
\nabla^2 f &=&  m_S^2 f  +f \left[ e^2 Q^2   -{\lambda \over 2} f
\left( 3 \eta - f
\right)  \right] -e^2 \eta Q^2  \\
\nabla \times \left( \nabla \times  Q \right) &=&  m_A^2 Q  - e^2( 2 \eta f -
f^2) Q,
\end{eqnarray}
\noindent where $m_S^2= \eta^2 \lambda$, $m_A=\eta e$ and $Q =
A_\theta -{n \over e r}$.

Equation (6) has an obvious solution for small $f$, when the second
term, on the right hand side, can be neglected in face of the
first\footnote{This is a special case of the most general solution
$A_\theta(r) = {n \over e r} - k_A K_1(e \varphi r)$, when $\varphi
\simeq$ constant.}  \cite{Nie}
\begin{equation}
A_\theta(r) = {n \over e r} - k_A K_1(m_A r),
\end{equation}
\noindent where $k_A$ is a constant and $K_1(m_A r)$ the modified
Bessel function of order 1. Equation (7) shows us how $A_\theta$ approaches
its asymptotic value. This result allows us to estimate the
corresponding behavior for the scalar field. Neglecting the quadratic
and cubic terms on the fields in (5) we obtain
\begin{equation}
f=k_S K_0(m_S r),
\end{equation}
\noindent where $k_S$ is another constant and $K_0(m_S r)$ the
modified Bessel function of order 0. For large arguments the modified
Bessel functions of order 0 and 1 have the same leading behavior, namely,
\begin{equation}
K_n(m r ) \sim  \sqrt{ \pi \over 2 m r} e^{ - mr}   \left( 1 + O({1 \over m r})
\right).
\end{equation}

The approximate solutions (7) and (8) when taken in the limit (9)
indeed  reproduce the
exact asymptotic form of the fields at critical coupling \cite{Vega}.
As we have seen above, however, they are valid for
all values of the coupling constants in the model.

For short distances, when $(\eta - \varphi) \simeq \eta$, equation (6)
becomes essentially that for a free gauge field. The corresponding
form for $A_\theta$ is then
\begin{equation}
A_\theta(r) \propto {n \over e r } + O(r).
\end{equation}

This form is actually compatible with (7), when the small argument for
the Bessel function is assumed. Moreover, if we impose that the
magnetic flux should vanish at the origin we will determine
$k_A = n \eta$ \cite{Nie}. This is  certainly an acceptable
procedure for strongly type II vortices.
An approximate solution for the scalar field close to the origin can
equally be found by using (10) in (5). We obtain
\begin{equation}
\varphi (r) \propto J_n({m_S \over \sqrt{2}} r),
\end{equation}
\noindent where $J_n$ is the Bessel function of order $n$, which is
the usual winding number. This
implies the well-known small distance behavior for $\varphi$
\begin{equation}
\varphi (r) \propto {1 \over 2}\left({m_S \over \sqrt{2}} r \right)^n.
\end{equation}
In particular for $n=1$ the field is linear close to the
origin.
Because the behavior of the scalar field in these two quite different
limits is not directly relatable, $k_S$ cannot be computed with
generality.
In the rather special case of critical coupling (${e^2 \over
\lambda}=1$), however, the dynamical equations for the fields
reduce to
\begin{eqnarray}
\partial_r \varphi \pm (e A_\theta - {n \over r})&=& 0   \\
\partial_r A_\theta  \pm {\sqrt{\lambda} \over 2} (\varphi^2 - \eta^2)&=& 0,
\end{eqnarray}
\noindent where $\pm = \rm{sign}(n)$.
These are the Bogomol'nyi equations for critically coupled vortices
in polar coordinates. It is easy to see that they allow us to relate
$f$ and $A_\theta$ directly and consequently also $k_S$ and $k_A$.
Using the identity of the two coupling constants we obtain $k_S= \vert
n \vert \eta$. In principle there is no reason for this result to hold
in other regions  of parameter space other than the good agreement
of its consequences with the numerical results, as we shall see later.
Nevertheless unless otherwise stated we will assume it henceforth.

We are now in a position to derive  the interaction potential between two
string segments, under the assumption that the solution for the two
string field configuration can be successfully approximated by the
superposition ansatz \cite{Abr,Mul}

\begin{eqnarray}
\Phi(r,r_1,r_2) &=& {\phi(\vert r-r_1 \vert) \phi( \vert r - r_2 \vert
) \over \eta } \\
A_\theta(r,r_1,r_2) &=&  A_{\theta}(\vert r-r_1 \vert) +
A_{\theta}(\vert r-r_2 \vert).
\end{eqnarray}

Here $\phi$ and $A_{\theta}$ are the isolated
string field
configurations of Section 1. In order to estimate the interaction
energy between the two vortices one then simply substitutes (15) and
(16) in the energy functional
\begin{equation}
E[\phi,A_\mu] =  \int dV \left[ {1 \over 4} F_{\mu\nu} F^{\mu\nu}
+ {1 \over 2} \vert\partial_r \phi \vert^2 + {1 \over 2}
 \vert e A_\theta \phi - \partial_\theta \phi\vert^2 +
{\lambda \over 8} ( \vert \phi \vert ^2 - \eta^2 )^2 \right].
\end{equation}
We then obtain

\begin{eqnarray}
E[\Phi,A] &=& E[\phi_1,A_1] + E[\phi_2,A_2] + \int dV {1 \over
2} \left(  2 \nabla_{r_1} \times A_{\theta}(r_1) . \nabla_{r_2}
\times A_{\theta}(r_2) +  \nonumber \right. \\
&+& {1 \over \eta^2} \left[ (\varphi(r_1)^2 -\eta^2) (\nabla_{r_2}
\varphi(r_2))^2 +
(\varphi(r_2)^2 -\eta^2)
(\nabla_{r_1} \varphi(r_1))^2 + \nonumber\right. \\
&+& 2 \varphi(r_1) \varphi(r_2)  \nabla_{r_1} \varphi(r_1). \nabla_{r_2}
\varphi(r_2) +
\nonumber \\
&+& (\varphi(r_2)^2 -\eta^2) \varphi(r_1)^2 (e A_{\theta}(r_1)
-{n_1 \over r_1})^2 +
(\varphi(r_1)^2 -\eta^2)
 \varphi(r_2)^2 (e A_{\theta}(r_2) -{n_2 \over r_2})^2 +
\nonumber \\ &+& \left. 2
\varphi(r_1)^2 \varphi(r_2)^2
(e A_{\theta}(r_1) - {n_1 \over r_1}). (e A_{\theta}(r_2)
- {n_2 \over r_2})\right] +
\nonumber \\ &+&
\left. {\lambda \over 4 \eta^4}\left[ (\varphi(r_1)^2 -\eta^2) (
\varphi(r_2)^2 -\eta^2) \left[ (\varphi(r_1)^2 +\eta^2) (
\varphi(r_2)^2 +\eta^2) - 2\eta^4 \right] \right] \right).
\end{eqnarray}
Finally, we subtract the contributions due to the isolated vortices. We
obtain
\begin{eqnarray}
E_{\rm int}[\varphi,A_\theta] &=& \int dV {1 \over
2} \left(  2 \nabla_{r_1} \times A_{\theta}(r_1) . \nabla_{r_2}
\times A_{\theta}(r_2) +  \nonumber \right. \\
&+& {1 \over \eta^2} \left[ (\varphi(r_1)^2 -\eta^2) (\nabla_{r_2}
\varphi(r_2))^2 +
(\varphi(r_2)^2 -\eta^2)
(\nabla_{r_1} \varphi(r_1))^2 + \nonumber\right. \\
&+& 2 \varphi(r_1) \varphi(r_2)  \nabla_{r_1} \varphi(r_1). \nabla_{r_2}
\varphi(r_2) +
\nonumber \\
&+& (\varphi(r_2)^2 -\eta^2) \varphi(r_1)^2 (e A_{\theta}(r_1)
-{n_1 \over r_1})^2 +
(\varphi(r_1)^2 -\eta^2)
 \varphi(r_2)^2 (e A_{\theta}(r_2) -{n_2 \over r_2})^2 +
\nonumber \\ &+& \left. 2
\varphi(r_1)^2 \varphi(r_2)^2
(e A_{\theta}(r_1) - {n_1 \over r_1}). (e A_{\theta}(r_2)
- {n_2 \over r_2})\right] +
\nonumber \\ &+&
\left. {\lambda \over 4 \eta^4}\left[ (\varphi(r_1)^2 -\eta^2) (
\varphi(r_2)^2 -\eta^2) \left[ (\varphi(r_1)^2 +\eta^2) (
\varphi(r_2)^2 +\eta^2) - 2\eta^4 \right] \right] \right).
\end{eqnarray}

This last expression can also be written in terms of the auxiliary
field $f$. Assuming cylindrical symmetry the energy per unit
length of the strings becomes

\begin{eqnarray}
E_{\rm int} &=& \int dS \left( H_1 H_2 \vec{{e_z}_1}.\vec{{e_z}_2} +
m_A^2 Q_1 Q_2 \vec{{e_\theta}_1}.\vec{{e_\theta}_2} \right) -
\nonumber \\
&-& \left( \nabla_r f_1 \nabla_r f_2 \vec{{e_r}_1}.\vec{{e_r}_2} +
m_S^2 f_1 f_2  \right) \nonumber \\
&-& {e^2 \over 2 } \left[ f_1^2 Q_2^2 +  2f_1^2 Q_1.Q_2 +  f_2^2 Q_1^2
+  2f_2^2
Q_1.Q_2 + 2 f_1 f_2 (Q_1^2 +Q_2^2) +  4f_1f_2 Q_1.Q_2 \right]
\nonumber \\
&+& \lambda ( 3 f_1^2f_2 + 3 f_2^2f_1) \left[ \eta -{3\over 8}(f_1 +
f_2) \right] +
{3\over 8}\lambda (f_1^3 f_2 +f_2^3 f_1).
\end{eqnarray}

To obtain the detailed behavior for the interaction energy one then
simply needs to perform the integrations in (19) or (20) assuming given
functional forms for the fields.
This procedure is absolutely straightforward but a few cautionary
remarks are in order.

Firstly, unlike the case of critical coupling, the proof for
the existence and uniqueness of multi-vortex solutions for general
couplings does not exist. In particular, any such solution will
not be static.
However it is known, from experimental
evidence in type II superconductors that multi-vortex configurations
indeed do occur.

Secondly the above procedure is clearly not exact. The superposition
ansatz effectively is not an exact solution of the Euler-Lagrange
equations (3) and (4). The resulting discrepancy on the value of the
interaction energy can nevertheless be computed. This can be achieved
by assuming residual
additive terms to (15) and (16) such that the resulting fields would solve
(3) and (4) exactly. Their behavior and contribution to the energy can
then be estimated whenever the functional form for the individual
vortex solutions is known \cite{Mul}. In particular for large
distances away from the strings' axes, when the modified Bessel
function behavior for $f$ and $Q$ applies, such contribution is  of
the order $e^{-2d} \over d$ \cite{Mul}. This is a second order subleading
effect
as we shall verify in the next sections.
This whole procedure allows us consequently not only  to state the
validity of the
superposition ansatz for a given functional form of the fields but
also to compute the magnitude of the approximation involved.

Finally, the integrations in (19) or (20) necessary to obtain the total
energy can, in general, also only be performed approximately. This will be the
subject of the two ensuing sections where expressions for the
interaction energy are indeed obtained, for all values of $b={e^2
\over \lambda}$ and arbitrary orientation of the two string
elements.

\section{The interaction energy between parallel strings as a function
of $b$ }

In this section we derive the integrated expressions for the
interaction energy, for all values of $b ={e^2 \over \lambda}$. For the
sake of simplicity we will restrict ourselves to the case of two
parallel strings and leave the study of arbitrary geometries to the
next section. The problem then reduces to the study of two coplanar
vortices. Because different values of $b$ imply quite
distinct field configurations we will analyze the cases of critical
coupling ($b=1$), type II ($b<1$) and type I vortices ($b
>1$) separately.

\subsection{The case $b=1$}

In view of the approximate field solutions obtained in Section 1 and
the expressions for the interaction energy (19) and (20) derived in
Section 2, this  is the simplest case to approach.
In general there are two length scales associated with each vortex.
They are simply the inverse classical masses for the gauge and scalar fields
i.e., $r_A=m_A^{-1}$ and $r_S=m_S^{-1}$, respectively. When the scalar
field acquires a non-zero expectation value ($ <\varphi>
\rightarrow_{r \rightarrow \infty} \eta$) both fields become massive
and the modified Bessel function solutions for the fields constitute a
good approximation. This should happen for distances $R$, measured
from the zero of the scalar field, such that $R>r_S$.

These approximate solutions tell us how the fields approach their
asymptotic values, which in turn correspond to the vacuum of the
theory. At distances $R$ larger than $r_S$ a vortex only perturbs the
vacuum by effectively acting as a source for the fields $Q$ and $f$.
The interaction  between two vortices, separated by a distance
$d>2 r_S$ then reduces, in the first approximation, to the interaction
between these two fields. This picture can indeed be obtained from (20)
if we  keep only the terms linear on the fields of each vortex. Then we
have
\begin{eqnarray}
E_{\rm int} &=& \eta ^2 \int dS n_1 n_2\left( m_A^2 K_0(m_A r_1)
K_0(m_A r_2)\vec{e_{z_1}}.\vec{e_{z_2}}+  m_A^2 K_1(m_A r_1)  K_1(m_A r_2)
\vec{e_{\theta_1}}.\vec{e_{\theta_2}}  \right) - \nonumber \\
&-& \vert n_1 \vert \vert n_2 \vert \left( \nabla_{r_1} K_0(m_S
r_1) \nabla_{r_2} K_0(m_S r_2) \vec{e_{r_1}}.\vec{e_{r_2}} + m_S^2 K_0(m_S r_1)
K_0(m_S r_2) \right).
\end{eqnarray}

Higher order terms will be much smaller due to the fact that the
fields themselves are exponentially decreasing. Moreover, their
contribution is of the order of the approximation involved in adopting
the superposition ansatz for the two-string field configuration.

To integrate (21) we first note that $\vec{e_{z_1}}.\vec{e_{z_2}}$ is
independent of the point on the plane. Then, using the fact that the
modified Bessel functions of order zero satisfy
\begin{equation}
\left(\nabla^2 -m^2\right)K_0(mr)=-2\pi \delta(r),
\end{equation}
\noindent in two dimensions, we can write
\begin{eqnarray}
E_{\rm int} &=&  2 \pi \eta ^2 \left[ n_1 n_2  K_0(m_A d) -
\vert n_1 \vert \vert n_2 \vert K_0(m_S d) \right] + \nonumber  \\
&+& \left[ \int dS  n_1 n_2 \left(K_0(m_A r_2) \nabla_{r_1}^2 K_0( m_A r_1)
+ m_A^2 K_1(m_A r_1)  K_1(m_A r_2)\vec{e_{\theta_1}}.\vec{e_{\theta_2}}
\right) - \right. \nonumber \\
&-& \left. \vert n_1 \vert \vert n_2 \vert \left(  K_0(m_S r_2) \nabla^2_{r_1}
K_0( m_S r_1) + \nabla_{r_1} K_0(m_S
r_1) \nabla_{r_2} K_0(m_S r_2) \vec{e_{r_1}}.\vec{e_{r_2}} \right) \right].
\end{eqnarray}
Now by noting that
\begin{equation}
\nabla_r K_0(mr) = - m K_1(mr); \qquad \nabla_{r_1} K_0(mr_2) = - m {r_1 -
d \cos( \theta_1) \over r_2} K_1(mr_2),
\end{equation}
and that
\begin{equation}
\vec{e_{\theta_1}}.\vec{e_{\theta_2}} = \vec{e_{r_1}}.\vec{e_{r_2}}= {r_1 -
d \cos \theta_1 \over r_2},
\end{equation}
we see that the integral in (23) vanishes and the interaction
energy becomes exactly\cite{Bat}
\begin{equation}
E_{\rm int}(d) = 2 \pi \eta^2  \left( n_1 n_2  K_0(m_A d) - \vert n_1 n_2 \vert
K_0(m_s d)
\right).
\end{equation}

This shows in particular that, in the limit of our approximations,
two vortices at critical coupling are free. This is equally true for
a pair of parallel cosmic strings, in agreement with well-known
numerical results \cite{Jac}.

\subsection{The case $b<1$}

Away from critical coupling the two length scales $r_S$ and $r_A$
become different and the fields are expected to change accordingly.
In the specific case when $e^2<\lambda$, $r_A$ will be larger than
$r_S$.
These two lengths now define a set of three coaxial cylinders,
around the axis of the string, where  the fields may behave in quite
a different fashion. Essentially, for distances $R$ larger than $r_A$
or smaller than $r_S$ the picture of the last subsection remains
qualitatively unchanged.
Indeed, for $R<r_S$ the fields must vary substantially from their
values at the origin to a value not too far from their asymptotic
forms.

For $R>r_A$, on the other hand, the scalar field should be very close
to its asymptotic value and consequently will be a slowly varying function of
$r$. Then both  fields will have acquired their
classical masses. For this case the computations of subsection 4.1 are
certainly valid and it follows that two vortices will repel and a pair
vortex-antivortex will attract, with magnitude given by (26).

The behavior in the intermediate region, however, turns out to be the
most interesting.
There $R>r_S$ and the fields are expected to acquire mass. This
implies, namely, that the modified Bessel function form for $Q$ holds.
However because $m_A R<1$, the Bessel function can be approximated by
its small argument behavior
\begin{equation}
Q= - n \eta K_1(m_A R) \simeq - {n \over e R}.
\end{equation}
When this form is used in the Euler-Lagrange equation for the scalar
field, assuming $\varphi = \eta -f$, we obtain

\begin{equation}
\phi \simeq 1- {n^2\over (m_S r)^2} +O({1\over (m_S r)^4}),
\end{equation}
\noindent which is the solution for the global string, in the same
regime ($R>r_S$)\cite{She}.

It should now be clear that, as we take $e^2 \rightarrow 0$, this
intermediate region grows to fill all space (from $R=r_S$ to
infinity), and the global string behavior is fully recovered.
This is not very surprising, of course,  since in this limit the gauge
field decouples from the scalar field and the Euler-Lagrange equation
for the latter is that for the theory exhibiting the global symmetry
alone.
What is nevertheless quite interesting is that for an ensemble of type
II local strings to interact as in the global case they simply need to
lie at relative distances smaller than $r_A$.{\footnote{We should keep
in mind that this result is a direct consequence of the value taken
for $k_A$, in Section 1.}}

Let us now see in detail how this picture arises from the previous
expressions for the interaction energy.
The part of the energy  dependent on $f$ can be integrated to give
a contribution which falls with at least ${1\over m^4_S d^3}$.
Because $m_S d>1$ this contribution will be small. On the other hand
since the modified Bessel function of order 1 is still a solution for
the gauge field the computations of subsection 4.1 referring to it
will still hold. We then obtain, keeping the terms on the gauge field
only
\begin{equation}
E(d) \simeq 2 \pi \eta^2 n_1 n_2  K_0(m_A d)\simeq - 2 \pi \eta^2 n_1 n_2
\ln(m_A d),
\end{equation}
\noindent since $m_A R$ is small for $R<r_A$.
However, our assumptions about the nature of the fields leading to (29)
were only valid for $R>r_S$, whereas in (26) we implicitly took it to be
applicable in all space between the origins of each vortex.
We must, therefore subtract the energy corresponding to $R<r_S$.
We then get
\begin{equation}
E(d) \simeq - 2 \pi\eta^2 n_1 n_2
\left( \ln(m_A d) -\ln(m_A r_S) \right) =  -2 \pi \eta^2 n_1 n_2 \ln(m_S d).
\end{equation}
This expression is just what we would expect for the global string,
c.f. \cite{She}, and ensures the consistency of
the arguments above. As is well known this results in  a strong,
infinite range\footnote{ In the case of nonvanishing $e$, $m_A^{-1}$
provides the natural cutoff at large $r$}, repulsive
(attractive) force for a pair vortex-vortex (vortex-antivortex).

\subsection{The case $b>1$}

In the preceding analysis we were able to compute estimates for the
interaction energy between vortices under a few quite general
assumptions. In particular we learned that the proximity of the scalar
field to its asymptotic value (traduced in the criterion $R>r_S$)
introduced a fundamental qualitative change in the behavior of the
fields, allowing us to use the approximate solutions (7) and (8) with
some confidence. The case of type I vortices is naturally
characterized by the fact that $r_A<r_S$.
The results for $b=1$ for the interaction between two vortices
should in particular still be valid as long as their scalar cores do
not overlap, i.e. for  ($d>2 r_S$). Then, unlike what happened when $b<1$,
and because now $m_A>m_S$ the scalar field contribution will dominate
the energy. Moreover,
 since the corresponding term in the interaction energy (26)
has no dependence on the sign of the winding numbers this
will always result on an attractive force regardless of the actual
nature of the two vortices involved.

As may easily be anticipated we will run into a problem when we try to
estimate what happens in the intermediate region ($r_A<d<r_S$). This can be
explicitly shown if
we search for the behavior of the fields of a single vortex in
this region of space.

 In fact, the way we derived the Bessel
Function behavior of $Q$ for large distances was to assume that
$\varphi$ would be close to $\eta$, so that the mass term in (6) would
clearly be the dominant one. This always holds provided that
$R \stackrel{>}{\sim} r_S $, and  is thus certainly a good approximation for
the cases above.

If, on the other hand, $R$ is small enough so that
$(\varphi^2 -\eta ^2)\simeq -\eta^2$,
and $Q$ exhibits its short distance behavior, then the solution
of (3) becomes
\begin{equation}
\varphi(r) \propto J_n({m_S r \over \sqrt{2}}) \simeq {1 \over 2}
({m_S r \over \sqrt{2}})^n.
\end{equation}

But this is the behavior we expect for the scalar field very close to
the origin, independently of any particular value of the coupling constants.

However,  we also note that as we progress towards $R=r_S$, $f$
approaches $\eta$ and the Bessel function behavior (7), and (8) start
being valid for $Q$ and $f$, respectively. When this happens, $Q$
automatically assumes an exponentially decreasing form since $m_A R$
is large but $f$ should still behave approximately logarithmically
since $m_S R$ is small. This should happen in a region
where the distance to the string's axis is sufficiently
close to $r_S$. Consequently and unlike what happened for type II
strings  this approximate behavior for the fields can only exist for
a thin region of space around a string. Its consequences for the
interaction energy between the two strings will then only constitute a
transient regime between its form for large distances and the actual
superposition of the two vortices.
Nevertheless we can compute what the corresponding
interaction energy should be.
Proceeding in the same way as in 4.2, but now only keeping the
contribution from the scalar field we obtain
\begin{equation}
E(d) \simeq - 2 \pi \eta^2 \vert n_1 n_2 \vert K_0(m_S d)\simeq  2
\pi \eta^2
\vert n_1 n_2 \vert \ln(m_S d),
\end{equation}
\noindent where, again, we implicitly assumed that this regime was
present in the whole
region between the two axes. We should then remove the contribution
from the inner region where this does not apply.
 In the absence of another length scale we
can only subtract the energy arising from $R<r_A$. This however will
result in an overestimate (in absolute value) for the final result. We
then obtain
\begin{equation}
E(R) \simeq  2 \pi \eta^2 \vert n_1 n_2 \vert \ln(m_A d).
\end{equation}
This  always gives rise to an attractive force, as expected.

At distances smaller than $r_S$, the scalar cores of the two strings
will superimpose. This results in a field configuration of winding
number $n_1+n_2$ as the distance between the zeros of the scalar
fields vanishes. Such a transformation of the two string field
configuration is accompanied by the change in behavior between the
modified and unmodified Bessel functions solution for the scalar field
of section 1. This change is
dictated by the nonlinearities in the corresponding Euler-Lagrange
equations which implies the breakdown of the superposition ansatz.
Close to $d=0$, however it is known numerically \cite{Jac} that the interaction
energy should become approximately constant, signalling the fact that a
type I string of higher winding number is a {\it stable} solution
of the Euler-Lagrange equations relative to its lower winding number
isolated constituents. [Alternatively, type II strings of higher
winding number are unstable with respect to decay into strings of
lower winding number].

\section{ Interaction between two cosmic strings: general geometry}

In the previous section we analyzed how the value of the ratio of the two
coupling constants in the model $b$ changes the field
configurations and derived the interaction energy  between two vortices,
i.e. between a pair of parallel strings, per unit length.

For a general network of cosmic strings we know\cite{Ben}
that strings wiggle
and bend, possibly on several scales, and, in general, no two strings are
parallel\footnote{Unlike cosmic strings, flux lines in
superconductors and vortices in superfluids are indeed parallel. For
those cases the results of Section 4 apply.}. We shall expand upon
this in the next section.

The assumption of parallel strings, however, conveniently allowed us some
comfortable simplifications. Because each string is locally (i.e. for
any element of infinitesimal length along its axis) cylindrically
symmetric we could associate the natural frame for the second string
with that for the first  simply by translating it by a distance $d$,
within the same plane.

This distance was then automatically defined for all pairs of string
elements and
the interaction energy for a piece of string due to the presence of
the second is simply that for one element times the length.
For a general configuration of the two strings
both the distance and orientation have to be specified for each pair
of elements. They will therefore weigh differently in the evaluation
of the interaction energy between two pieces of string of finite
length.

Because of the cylindrical symmetry of the fields composing a string
is  still locally preserved, to find the interaction energy of one of
its elements we need only look for an element of a second string
intersecting the plane in which it lies. The natural frame for the
element of the second string will appear in the general case to have
its origin at a distance $d$  but also to have been rotated so that
it no longer lies in the plane of the former. However, because of its
own
local cylindrical symmetry, rotations around its axis of symmetry leave
it unchanged. This rotation relative to the plane of the first vortex
is then generated by two angles only.
This can be seen in figure 1, where we chose the x-axis along the
direction connecting the origins of the two vortices. The angles
$\alpha$ and $\gamma$ parameterize rotations around the $x$ and $y$-axes,
respectively.

In order to be able to generalize the results of Section 3 we must be
able to understand how the change in orientation of the two relative
elements affects the interaction energy. This will result essentially
in a change on the inner products between the unit vectors for the
directions associated with the natural frame of each string.

This change can be computed by rotating one of the frames around the x
and  y-axes of figure 1.
We now see, in particular, that $\vec{e_{z_1}}.\vec{e_{z_2}}$ will
simply have the form
\begin{equation}
\vec{e_{z_1}}.\vec{e_{z_2}}= \cos(\alpha) \cos(\gamma).
\end{equation}
This allow us to integrate the two terms in (23) to obtain
\begin{eqnarray}
E_{\rm int} &(& d,\alpha,\gamma) = 2 \pi \eta ^2 \left[ n_1 n_2  K_0(m_A d)
\cos(\alpha) \cos(\gamma)-
\vert n_1 \vert \vert n_2 \vert K_0(m_S d) \right]  - \nonumber \\
&-& \left[ \int dS  n_1 n_2 \left( \nabla_{r_1} K_0(m_A r_2)
\nabla_{r_1} K_0( m_A r_1)
\cos(\alpha) \cos(\gamma)
- K_1(m_A r_1)  K_1(m_A r_2)\vec{e_{\theta_1}}.\vec{e_{\theta_2}}
\nonumber \right) \right.\\
&-& \left. \vert n_1 \vert \vert n_2 \vert \left( \nabla_{r_1} K_0(m_S
r_2)  \nabla_{r_1}
K_0( m_S r_1) - \nabla_{r_1} K_0(m_S
r_1) \nabla_{r_2} K_0(m_S r_2) \vec{e_{r_1}}.\vec{e_{r_2}} \right) \right].
\end{eqnarray}

The two residual terms, inside the integral, behave somewhat
differently. In the case of that arising from the scalar field the
change in variable of differentiation on the first exactly generates
the inner product $\vec{e_{r_1}}.\vec{e_{r_2}}$. Because both these terms
have opposite signs their sum vanishes.

Such is not the case for the corresponding term in the gauge sector.
There we have
\begin{eqnarray}
{dr_2 \over dr_1} \vec{e_{z_1}}.\vec{e_{z_2}} &=& {1 \over r_1 r_2}
\cos(\alpha) \cos(\gamma) \left[ x_1(x_1-d)
\cos^2(\gamma) + y_1^2 \cos^2(\alpha)  \nonumber \right. \\
&+&  x_1 z_1 \cos(\alpha) \cos(\gamma)
\sin(\gamma) - x_1 y_1 \sin(\alpha) \sin(\gamma) \cos(\gamma)
\nonumber \\
&+& \left. y_1z_1
sin(\alpha) \cos^2(\alpha) \right],
\end{eqnarray}

\noindent which must be compared with

\begin{eqnarray}
\vec{e_{\theta_1}}.\vec{e_{\theta_2}} &=& {1 \over r_1 r_2}  \left[
\cos(\alpha) \cos(\gamma) [ x_1(x_1-d)
 + y_1^2  ]  \nonumber \right. \\
&+& \left. x_1 z_1 \cos^2(\alpha) \sin(\gamma)
 - x_1 y_1 \sin(\alpha) \cos(\alpha) \sin(\gamma) + y_1z_1
sin(\alpha) \cos(\gamma)  \right].
\end{eqnarray}
In both the above expressions we took a rotation by $\gamma$ after a
rotation by $\alpha$.

These two terms do not cancel in general. Rewriting (35) we finally
obtain
\begin{eqnarray}
E_{\rm int}(d,\alpha,\gamma) &=& 2 \pi \eta ^2 \left[ n_1 n_2  K_0(m_A d)
\cos(\alpha) \cos(\gamma)-
\vert n_1 \vert \vert n_2 \vert K_0(m_S d) \right] - \nonumber  \\
&-& \left[ \int dS  n_1 n_2  \nabla_{r_2} K_0(m_A r_2)
\nabla_{r_1}  K_0( m_A r_1) \left( {1 \over r_1 r_2} \left[ x_1(x_1-d)
\sin^2(\gamma) + \right. \right. \right.
\nonumber \\ &+& \left. \left. \left. y_1^2 \sin^2(\alpha)
+  {\rm higher \ order \ in \ \alpha, \gamma} \right] \right)   \right]
\cos(\alpha) \cos(\gamma).
\end{eqnarray}

We see that the interaction energy is essentially given by the two
first terms. In the limit of vanishing $\alpha$ and $\gamma$ the
residual contribution in the integral effectively goes to zero with
the sine squared of the angles, in agreement with the results of Section 3.

Expression (38) stresses the  quite different nature of the two
contributions to the interaction energy. The vector-like character of
the gauge field introduces a dependence on the relative orientation
between the two strings. In particular we see that if one of the
strings is rotated by $\pi$ the corresponding interaction energy term
changes sign. This
is equivalent to changing the nature of a vortex into an anti-vortex
or vice-versa and results from the geometric nature of the winding
number.
The term arising from the scalar field is, in contrast, insensitive to
any  particular configuration for the two-string system, as could
naturally be anticipated.

It also becomes clear that critical coupling ($b=1$), for an arbitrary
geometric configuration, ceases to be the special case when the
interactions between strings vanish since the
angular dependence on $\alpha$ and $\gamma$ destroys  the balance
between the gauge and scalar terms in the energy.

All expressions derived above concern elements of string with
infinitesimal length, resulting from the integration of the
interaction energy functional on the plane. To obtain the interaction
energy for two segments of string of finite length one then has to
integrate over a series of planes perpendicular to one of the strings.
In particular, every time the string bends so that any of its
elements will lie at an angle larger than $\pi \over 2$  it is
expected to experience self-interactions.
This is always the case for closed loops of string which
contain part of the energy arising from self-interactions, in
addition to their tension\footnote{If the string's tension is defined to be the
energy/unit length for an infinite straight string}.

The expressions derived above allow us to compute analytically the
interactions for a variety of configurations. Let us consider two
simple examples.

For a circular loop of large enough diameter $D$ ($D>2r_S$ as always)
it is straightforward to compute its self-energy. It is simply
\begin{equation}
E_{self}(D) = {\pi D \over 2} E(D),
\end{equation}
\noindent where $E(D)$ is given by  expression (26), taken in the
appropriate limit, for a
parallel pair of string-antistring ($ \vert n_1 \vert =-\vert n_2 \vert$).

Another interesting case is to compute the interaction energy due to a
circular loop of string, in the limit when the distance to a test string
segment is much larger than the loop's diameter. Then, see figure 2,
the interactions are essentially due to a dipole of string segments.
The corresponding interaction energy is proportional to the sum of the
Bessel functions coming from the gauge and scalar field contributions.
In particular for the contribution from the former we will have
\begin{equation}
E_{\rm int} \propto \left( K_0(m r) - K_0(mr_2) \right),
\end{equation}
\noindent where
\begin{equation}
r_2=\sqrt{D^2 +r^2 +2Dr \cos(\theta)}.
\end{equation}
Now, in the limit of small diameter compared to distance ${D \over
r}<<1$ and for the strong type II  small loops of string, when
$D m_A \cos(\theta)<1$,
we can series expand the Bessel function to obtain
\begin{eqnarray}
K_0 \left( mr \sqrt{1+  {D^2 \over r^2} + 2 {D \over r} \cos(\theta)}
\right) &\simeq&
K_0(mr) \left[ 1+ m^2 D^2 \cos(\theta) \right] - \nonumber \\
&-& mD \left[ \cos(\theta)-
{D \over r}  \sin^2(\theta) K_1(mr)\right].
\end{eqnarray}

The energy can then  approximately be written as\footnote{Here we
have neglected the contribution form the scalar field and took the
winding number of the test string segment to be one. It is also
assumed to lie in a plane parallel to that of the loop.}
\begin{equation}
E_{\rm int}\simeq 2 \pi \eta^2 n \left( mD \left[ \cos(\theta) - { D
\over r} \sin^2(\theta) \right]
K_1(mr) - m^2 D^2 \cos^2 (\theta) K_0(mr) \right).
\end{equation}
This expression has two different limits depending on the relative
value of $m_A^{-1}$ and $r$. If $m_A r$ is large then we can take the
asymptotic forms for the Bessel functions at large arguments (9) to obtain
\begin{equation}
E_{\rm int} \simeq 2 \pi \eta^2  \sqrt{ \pi \over2 m_A r} e^{- m_A r}  \left(
m_AD
\cos(\theta) - \sin^2(\theta){D \over r} -m_A^2 D^2 \cos^2 (\theta)\right).
\end{equation}
If, on the other hand, $r<m^{-1}_A$ ( extremely type II
case), we can use the form of the Bessel functions for small
arguments to obtain
\begin{equation}
E_{\rm int} \simeq 2 \pi \eta ^2 \left[ m_A D \left( \cos(\theta) - {D \over r}
\sin^2(\theta)
\right) {1 \over m_A r} + m_A^2 D^2 \cos^2(\theta) \log(m_A r) \right].
\end{equation}
We see in particular that in this latter case the leading order term
in (45)  coincides with the usual ${1 \over r}$ potential for a
point-like field source in 3 dimensions. A small loop of (quasi)global
string thus behaves effectively as a monopole when seen at
sufficiently large distances.

\section{String formation and string forces}

In simulations of string production (e.g., see \cite{Vach}) $\phi$ field
phases are laid down at random in a space divided into correlation
volumes within which the $\phi$ field phase is held approximately constant. In
these
circumstances strings with winding number $\vert n\vert =1$ are produced at
high density. The laboratory
production \cite{Uoojy} of string defects in superfluid $^4 He$ has suggested
that this
model is plausible \cite{Zurek}, at least for the non-relativistic
vortices of the global $U(1)$ Ginzberg-Landau theory.

In addition to the role string interactions have in determining the
outcome of string collisions  they may
affect the early string evolution if, as expected,
strings are produced at sufficiently high density. In the remainder of
this paper we shall examine the effect of interstring forces in a newly
created, and approximately static, ensemble of strings.  In this section
we shall investigate whether, with
only attractive forces, type I strings can exist at high density. This
is not a question of how two strings of lower winding number combine
to form a string with higher winding number, since we expect strings,
of whatever type, to appear as a tangled mess.
However, we shall return to this point later.
To see how type I
strings could exist we adopt a classical toy model possessing some of
the characteristics of real strings, in which the effect of attractive
forces is more transparent.

Cosmic strings are
produced at phase transitions in the G.U.T. era but, nominally independently
of this, it
is known that thermodynamic ensembles of classical string naturally
display transitions. Most naively, if cosmic strings are treated as
non-interacting Nambu-Goto strings with modes of arbitrarily high
frequency they undergo a Hagedorn transition \cite{Neil}.In this section
we take this classical transition seriously, but we shall be more
realistic by treating classical strings more as polymers. These again
show transitions but, unlike the fundamental Nambu-Goto strings,
permit the inclusion of string forces at low densities.

The comments that follow are a straightforward extension of earlier
work of ours on global strings \cite{Cope}. Since they are somewhat
speculative a detailed recreation of these earlier results is
inappropriate, and the reader is referred to the literature.

 Strings
produced in the way indicated above behave like random walks of high
density \cite{Scherrer}. It is
well-known that an ensemble of non-interacting random walks at
temperature $T=\beta^{-1}$, step length $a$, can be described by a
dual field theory of a free complex field $\chi(\underline x)$
\cite{Thomas}. The ``loop field'' $\chi(\underline x)$ has action
(Hamiltonian) \cite{Thomas}
\begin{equation}
S_0 = \int d \underline x \left[ \vert \nabla \chi \vert ^2 + M^2
\vert \chi \vert ^2 \right],
\end{equation}
\noindent where
\begin{equation}
M^2 \propto { \left( \epsilon - Ts \right) \over a T },
\end{equation}

\noindent in which $\epsilon$ is the string energy/unit length and
$s= O(a^{-1})$ is the entropy/unit length. The vanishing of $M^2$ at
$\epsilon = Ts$ defines a Hagedorn transition at $T=T_H$, at which
string is produced copiously.  For $T>T_H$ the theory
is unstable\footnote{And at which all energy goes into the production
of a single string.}.

This is already enough to indicate why, even if the string formation
mechanism naturally generated strings with higher winding number
$\vert n\vert >1$, the thermal energy would be sufficient to unpeel them to
$\vert n\vert =1$
strings.
The problem is akin to that of adhesion in macromolecules.  As a first
step we adopt the simplifying assumption that, once separated from a string of
higher winding
number, $\vert n\vert =1$ strings experience no forces.  A
straightforward extension \cite{Sonya} of the work of Wiegel \cite{Wiegel}
shows that, provided the energy required to split an
$n=2$ string into two $n=1$ strings is less than the cost of creating
the string, the Hagedorn transition survives at the same temperature $T_H$.
Moreover, there is a temperature range $T_0 <T< T_H$ in which the
separation into $n=1$ strings is total.  We shall not consider the
problem of higher winding number further.

The forces between the strings modify (46). Orientable forces are
characterized by a dual vector field ${\cal A}$. It is not possible to mimic
our cosmic strings exactly. However, for a strong type II theory of
static strings at low density, a plausible dual field theory is given by an
action \cite{Cope}

\begin{equation}
S_{II} = \int d \underline x \left[ {1 \over 4} \left( \nabla \wedge
{\cal A}  \right)
+ {1 \over 2} m_A^2 {\cal A}^2 + \vert \left( \nabla + i f {\cal A}
\right)
\chi \vert^2 + M^2 \vert \chi \vert ^2 \right],
\end{equation}

\noindent where $m_A \simeq 0$, and $f^2 = O( \beta^2 \epsilon)$ is
the dimensionless coupling strength. $M^2$ is as before.

By the same token, the extension of $S_{II}$ to include a scalar field
$\sigma$, as in the action

\begin{eqnarray}
S = \int d \underline x \left[ {1 \over 4} \left( \nabla \wedge
{\cal A}  \right)
\right. &+& {1 \over 2} m_A^2 {\cal A}^2 + \vert \left( \nabla + i f {\cal A}
\right)  \chi
\vert^2 + \nonumber  \\
&+& \left. {1 \over 2} ( \nabla \sigma )^2 + {1 \over 2} m_S^2 \sigma^2
- g^2 \sigma^2
\vert \chi \vert ^2 + M^2 \vert \chi \vert ^2 \right],
\end{eqnarray}

\noindent  corresponds to the inclusion of non-orientable sources and
hence the possibility of type I strings. The coupling $g^2 > 0$
characterizes the attractive nature of these forces. The simple
diagonal form ( in terms of $\sigma$ and $\cal{A}$) cannot fully reproduce the
results of (38). However, it does indicate how attractive forces of
finite range are permissible at low density.

To see this, we note that, in the dual theory of (46), (48) or (49),
the dimensionful string density $\rho= {L \over V}$, the length/unit
volume, is given by $< \vert \chi \vert^2>$ \cite{Thomas}. We see
from (49) that, because of the attractive forces, the classical
$\sigma$-field sector is unstable unless $\vert \chi \vert ^2 < {m_S ^2
\over g^2}$, i.e., unless the string density is sufficiently low. We
note that long-range wholly-attractive forces would always lead to
(classical) instability in this simple model. However, the result  that finite
range attractive forces do not destabilize the network at low enough
density (large enough separation) seems very plausible. Further,
insofar as $g^2$ increases as the strings become more strongly type I,
the maximum permissible density drops, given the constancy of $m_S^2$.

To do better would require a greater confidence in (47) that we can
expect. However, a
one-loop approximation to the effective density potential in which the
$\cal{A}$ and $\sigma$ fields are integrated over in (49) gives

\begin{eqnarray}
V(\rho)= M^2 \rho &+& 3 T \int \not \! d \underline k \ln ( \underline k^2
+ m_A^2 + 2 f^2 \rho ) +\nonumber \\
&+&  T \int \not \! d \underline k \ln ( \underline k^2
+ m_S^2 - 2 g^2 \rho ).
\end{eqnarray}

$V(\rho)$
of (50) is sufficient to show why, when arbitrary string orientations
are taken into account, there is nothing special about the value
$b=1$.
For $b$ near unity, when $ m_S \simeq m_A $ and (presumably) $ f \simeq
g $, the effect of the attractive non-orientable third term is to
partially, but not wholly, cancel the effect of the orientable force
term at low density. This is as we would expect.

For strong type I strings the situation is totally unclear, and there
is no natural way to identify type I and type II strings with
first-order (melting) and second-order (Hagedorn) transitions,
directly. It could be argued that neither should there be, since
whether a string is type-I or type-II has more to say about the nature
of the quantum field phase transition than classical transitions. It
is to the quantum aspects of string production that we now turn.

\section{String formation from quantum fluctuations}

The main reason for halting our analysis of the dual theory is not so
much its incompleteness as the fact that, initially, cosmic string
production arises from quantum fluctuations at the phase transition
possessed by the underlying local $U(1)$ field theory. Without
reference to string, in approximate
thermal equilibrium, the theory displays a second-order transition at
the temperature $T_c = O(\eta)$ if $b << 1$. On the other hand, for
$b>>1$ it is reliably expected to display a first-order transition.
This accords with our terminology of type II and type I strings,
respectively. However, there is no reason to believe that the
transition changes from second to first order exactly at
$b=1$.

Whatever the case, in quantum field theory there cannot be a Hagedorn
transition, with its characteristic maximum temperature.
Rather, there is a critical temperature $T_c$ above which the $U(1)$
symmetry is restored and strings cannot occur. In thermal equilibrium
at temperature $T$, this is best seen by working in Euclidean time, in
which the fields are periodic with period $\beta=T^{-1}$. Provided $T$
is larger than the mass scales the 'heavy' ($n \neq 0$) modes in the
Fourier series can be integrated out, to give an effective
three-dimensional theory derived from (1), with action \cite{Haws,Kajantie}
\begin{eqnarray}
\beta S_3 &=& \int d \underline x \left[ {1 \over 4} \left( \underline
\nabla \wedge \underline A \right) + {1 \over 2} \vert  \left( \underline
\nabla + i e \underline A \right) \phi \vert ^2 - {1 \over 2} m^2(T)
\vert \phi \vert^2 + {\lambda \over 8} \vert \phi \vert ^4 \right]
\nonumber \\
&+&  {\rm terms\  containing\  A_0(\underline x)\ +\ counterterms},
\end{eqnarray}
\noindent where $A_0(\underline x)$ is the temporal component of the ``light''
($n=0$) mode  component of the gauge-$\underline{A}$ field, and

\begin{equation}
m^2(T) = {\lambda \eta^2 \over 2} \left( 1- {T^2 \over T_c^2} \right),
\end{equation}

\noindent is the effective scalar mass of the theory.

There are two points to note about the action (51).
Firstly, when $A_0 = 0$, it is extremised by the same (static) vortex
solutions as was the action based upon (1), but for the fact that the
Higgs mass is temperature dependent. The second is that the Higgs
potential in (51) shows a second-order transition, prior to
integrating over the gauge field $\underline A$, whatever the value of
$b$. A first order transition can only occur for $b>>1$ as a
consequence of gauge-field radiative corrections.

Returning to the first point, we could use the 'classical' solutions
to (51) as the basis of a dual field theory, as in the previous
section. This would now be a dual theory in the more usual sense of
the word, a rewriting of the original quantum theory in terms of its
excitations. The instability of this theory is, in the first instance,
characterized again  by the vanishing of
\begin{equation}
{\cal U} = \epsilon -T s ,
\end{equation}
\noindent except that $\epsilon$ and $a$, the step length, are now
expressed in terms of $m(T)$. With $\epsilon = O({m^2(T) \over
\lambda})$ and $s=O(m(T))$ the condition ${\cal U} =0$ becomes $ {\lambda T
\over m(T)} = O(1)$ \cite{Haws}, or
\begin{equation}
\left( 1 - { T^2 \over T_c^2} \right) = O(\lambda).
\end{equation}
This is the Ginzberg criterion for the onset of large fluctuations in
the vicinity of $T_c$, at which (51) becomes an unreliable basis for
calculation.
Thus, the Hagedorn transition of the previous section becomes subsumed
in the conventional phase transition for the quantum field and the
'classical' and quantum pictures are reconciled. Our
comments  on attractive forces enforcing a maximum density if they are
not to lead to unstabilities also survives. Moreover, although
$M^2(T)$ diminishes towards the transition, so does the scalar dual
field coupling strength $g^2$, which we assume to be $O(\epsilon)=
O(m^2(T))$.
Thus the maximum density does not diminish in the way we might have
expected from the classical picture.

Nonetheless, given that the string density is so high in the naive
Kibble mechanism \cite{Scherrer}, it is always possible that this
bound is breached.
For type I strings another possibility exists.  A first order transition
is characterised by the production of latent heat.  In the context of
dual string theory it will be manifest as a first order transition in
the string density $\rho$. A naive model for dislocation melting in
solids has been based upon action $S_{II}$, treated as a 'quantum'
theory ($\hbar = 1$) \cite{Edwards}. Since melting is, by definition,
first-order, the 'quantum' fluctuations of the $\cal{A}$-field in (48) can
indeed induce a first-order transition in the order parameter $\rho$
if the coupling $f^2$ of the orientable forces is strong enough
\footnote{ There is no contradiction in Type-II strings having a first
order transition once the strength of the string forces can be chosen
arbitrarily, and are not fixed by the underlying field theory.}. The
inclusion of non-orientable forces clouds the issue.
Such a transition is extremely difficult
to predict, requiring an analysis of the dual theory beyond the
mean-field approximation hinted at earlier.  Calculations have been
performed \cite{Sonya} that extend the work of \cite{Cope} to include the
temperature dependence of the mass $m(T)$ and the coupling strength
$f$ for strings (extreme type II) with short-range orientable
forces. They show that, should a first-order transition occur, it
too will be buried in the Ginzberg temperature range.  The effect of
this is that the string density can be very much lower (e.g. a
factor $10^{-3}$) than we would have had otherwise.  Such a low
density is unlikely to have any problems for stability.

As a final comment, it might be argued that the use of static
networks, even initially, is inappropriate.  However, the difference
between an ensemble of static strings and an ensemble of
relativistic strings, while present, is not as much as might be
thought.  For example, relativistic Nambu-Goto loops are described
in terms of right-moving and left-moving modes.  This doubling of
modes changes the power behaviour of the prefactor to the Hagedorn
exponential in the counting of states, in comparison to a static
ensemble\footnote{It is this power which determines critical
indices.}.  However, this change is
exactly cancelled by integrating over the loop centre-of-mass
momenta, which effectively makes their centres of mass static.
In the same sense, the time-independent action (51), if used as a
basis for string saddlepoints, would give rise to zero-frequency
modes that could be interpreted as centre-of-mass loop dynamics.

Of course, we have been very simplistic  in our neglect of
quantum fluctuations about the strings. This has been considered
elsewhere (last reference, \cite{Cope}) and does not derail our
conclusions to date.  However, it does compound our inability
to be quantitative with the dual theory.

\section{Conclusions}

 We have shown how the interaction energy for $U(1)$ cosmic strings
depends on the parameter space of the model as well as on the relative
orientations of two interacting strings.

One of the crucial assumptions in all simulations of the evolution of
networks of cosmic strings is that they intercommute in all cases.
This property, however, should depend both on the topology of the
problem which ensures
the exchange of ends of the two strings and is model independent, but
also on the dynamics of the system under the effect of its
interactions.

Our analysis  reveals that type I strings experience
attractive and, in the limit where the contribution from the gauge
sector is negligible, non-orientable interactions. This is corroborated by
several numerical studies and implies that in the absence of a strong
effect at the superposition of their scalar cores two such strings
will form a bound state of higher winding number, at a sufficiently low
energy collision and for small relative angles. Such higher winding
number states are  well-known to be stable. A more detailed study
concerning this problem is presented elsewhere \cite{Bett}, where we
attempt to quantify the circumstances in which such configurations may occur.

Type I strings are especially interesting for scenarios of thermal
string production after a period of inflation,
since the critical temperature associated with the phase transition is
lowered by taking $b>>1$ \cite{Hod}. If the intercommuting of type I
strings could be shown to be very inefficient such scenarios would
have to be ruled out on the grounds of being cosmologically
unacceptable. The most likely effect of the interactions, however,
would probably be that of
modifying the evolution parameters of a network of strings to some
extent without jeopardizing the approach to a scaling regime \cite{Bett}.

Type II, global and critically coupled strings in our picture would
intercommute.
This is absolutely consistent with several numerical studies\cite{She,Mat}.

Finally the knowledge of explicit forms for the interaction energy of
strings is a crucial element for the construction of a realistic
statistical description of strings, as well as of vortices away from
critical coupling. In the last sections we have indicated, through the
methods of a dual field theory, how strings with attractive forces do
not destabilize the initial string network once the density becomes
low enough. Observed originally for classical strings, it follows
equally for strings produced by equilibrium fluctuations. Further, we
see why $b=1$ ceases to be  critical once arbitrary string
orientations are taken into account. Of course, strings produced by
fluctuations have to be frozen in by subsequent out-of-equilibrium
development, and this has been omitted from our discussion.
The whole mechanism of string production from a quantum
theory, both with and without initial approximate thermal
equilibrium, is considered elsewhere \cite{Ray}. Nonetheless, the
mechanisms proposed by Kibble \cite{Kib} for string production from
fluctuations, from which we have been quoting, seem substantially
correct.

\section*{Acknowledgements}

We would like to thank  Andy Albrecht, Ed
Copeland, Tom Kibble and Paul Shellard for having shared some of
their wisdom on this subject with us. L.M.A.B's research
was supported by J.N.I.C.T. under contract BD 2243/RM/92.

\clearpage
\section*{Figure Captions}


{\bf Figure 1} : A schematic example of a two-string configuration
and of the relevant
quantities necessary to compute the corresponding interaction energy.

\vspace {.1in}

{\bf Figure 2} : The field as measured by an observer away from a circular
string loop.

\end{document}